\documentclass[final,a4paper]{article}

\usepackage{url}

\usepackage[utf8]{inputenc}

\usepackage{tabularx}
\usepackage{paralist}
\usepackage{mdwlist}

\usepackage{afterpage}

\usepackage{amsmath,amssymb,amsthm}
\usepackage{xspace}
\usepackage{mathpartir}

\usepackage{float}
\restylefloat{figure}

\renewcommand{\implies}[0]{\mathrel{\Rightarrow}}

\DeclareMathOperator{\dom}{dom}

\newcommand{\Trm}[1]{\textrm{#1}}

\newenvironment{myfigure}[1][phtb]
{\begin{figure}[#1]
  \hrule\medskip\begin{center}}
{\end{center}\vspace{-.2em}\hrule\end{figure}}

\newsavebox{\lsavedbox}

\def\thmref{section}

\newtheorem{theorem}{Theorem}[\thmref]
\newtheorem{lemma}{Lemma}[\thmref]
\newtheorem{corollary}[lemma]{Corollary}
\theoremstyle{plain}
\newtheorem{definition}[lemma]{Definition}

\def\bottom{\perp}

\def\ra{\rightarrow}
\def\lra{\leftrightarrow}

\def\vertrel{\mathrel{\vert}}

\newcommand{\CCNAT}[0]{\ensuremath{{\texttt{\textnormal{CC}}}_{\mathbb{N}}}\xspace}

\newcommand{\rw}[1]{\mathrel{\rightarrow_{#1}}}
\newcommand{\rwequiv}[1]{\mathrel{\leftrightarrow_{#1}^*}}

\newcommand{\rwrefltrans}[1]{\mathrel{\rightarrow^*_{#1}}}

\DeclareMathOperator{\nat}{nat}

\def\Prop{\ensuremath{\star}}
\def\Type{\ensuremath{\square}}
\def\Extern{\ensuremath{\triangle}}

\def\Ar{\textnormal{\textbf{r}}}
\def\Au{\textnormal{\textbf{u}}}

\def\NO{\mathbf{0}}
\def\NS{\mathbf{S}}
\def\NP{\mathbin{\dot{+}}}

\newcommand{\Rec}[1]{\ensuremath{\textnormal{Rec}^{#1}_{\mathbb{N}}}}
\def\RecW{\Rec{\cW}}

\newcommand{\RecL}[1]{\ensuremath{\textnormal{Rec}^{#1}_{\mathbb{L}}}}
\def\RecLW{\RecL{\cW}}

\def\dlist{\mathbf{list}}
\def\Lnil{\mathbf{nil}}
\def\Lcons{\mathbf{cons}}
\DeclareMathOperator{\reverse}{reverse}

\DeclareMathOperator{\Eq}{Eq}

\DeclareMathOperator{\Elim}{Elim}
\DeclareMathOperator{\Ind}{Ind}

\def\lam[#1@#2#3]#4{[\lambda {#1} :^{#2} {#3}] {#4}}
\def\lamA[#1@#2]#3{[\lambda {#1} : {#2}] {#3}}

\def\pdt(#1@#2#3)#4{(\forall {#1} :^{#2} {#3}) {#4}}
\def\pdtA(#1@#2)#3{(\forall {#1} : {#2}) {#3}}

\def\elim(#1@#2[#3]->#4)(#5){\Elim({#1} : {#2}[#3] \rightarrow {#4})\{{#5}\}}

\def\ind(#1@#2)(#3){\Ind({#1} : {#2})\{{#3}\}}

\def\constr(#1,#2){{#1}^{[{#2}]}}

\DeclareMathOperator{\Class}{Class}

\newcommand{\ccieq}[0]{\mathrel{\dot{=}}}

\newcommand{\Alg}[0]{{\textnormal{\bfseries Alg}}\xspace}

\DeclareMathOperator{\fcap}{cap}

\newlength  {\CInferRuleWidth}
\newlength  {\CInferRuleWhenWidth}
\newsavebox {\CInferRuleBox}

\newcommand{\InferRule}[3]{\ensuremath{\inferrule*[left={[#1]}]{#2}{#3}}}

\DeclareMathOperator{\FV}{FV}

\newcommand{\cnv}[1]{\sim_{#1}}
\newcommand{\cnvbl}[1]{\cong_{#1}}

\newcommand{\cA}[0]{\mathcal{A}}

\newcommand{\cD}[0]{\mathcal{D}}
\newcommand{\cE}[0]{\mathcal{E}}

\newcommand{\cK}[0]{\mathcal{K}}

\newcommand{\cO}[0]{\mathcal{O}}
\newcommand{\cP}[0]{\mathcal{P}}

\newcommand{\cR}[0]{\mathcal{R}}
\newcommand{\cS}[0]{\mathcal{S}}
\newcommand{\cT}[0]{\mathcal{T}}

\newcommand{\cW}[0]{\mathcal{W}}
\newcommand{\cX}[0]{\mathcal{X}}
\newcommand{\cY}[0]{\mathcal{Y}}

\usepackage{url}

\title{%
  Building Decision Procedures in the\\
  Calculus of Inductive Constructions}

\author{Frédéric Blanqui\footnotemark[1],
Jean-Pierre Jouannaud\footnotemark[2] and
Pierre-Yves Strub\footnotemark[2]}

\newcommand{\authorsinfos}[0]{%
\footnotetext[1]{%
  LORIA, UMR 7503 CNRS-INPL-INRIA-Nancy2-UHP, Equipe Protheo,
  Campus Scientifique,
  BP 239, 54506 Vandoeuvre-lès-Nancy Cedex,
  \url{blanqui@loria.fr}}
\footnotetext[2]{%
  Projet LogiCal (Pôle Commun de Recherche en Informatique du
   Plateau de Saclay, CNRS, École Polytechnique, INRIA,
   Univ. Paris-Sud.),
  LIX, UMR CNRS 7161,
  École Polytechnique, 91128 Plaiseau, FRANCE,
  \url{{jouannaud,strub}@lix.polytechnique.fr}}%
}

\date{}

\begin{document}

\maketitle
\thispagestyle{empty}

\authorsinfos

\begin{abstract}
  It is commonly agreed that the success of future proof assistants
  will rely on their ability to incorporate computations within
  deduction in order to mimic the mathematician when replacing the
  proof of a proposition P by the proof of an equivalent proposition P'
  obtained from P thanks to possibly complex calculations.

  In this paper, we investigate a new version of the calculus of
  inductive constructions which incorporates arbitrary decision
  procedures into deduction via the conversion rule of the calculus.
  The novelty of the problem in the context of the calculus of
  inductive constructions lies in the fact that the computation
  mechanism varies along proof-checking: goals are sent to the
  decision procedure together with the set of user hypotheses
  available from the current context.  Our main result shows that this
  extension of the calculus of constructions does not compromise its
  main properties: confluence, subject reduction, strong normalization
  and consistency are all preserved.

  \medskip

\end{abstract}

\section{Introduction}
\label{s:intro}

\paragraph{Background.}
It is commonly agreed that the success of future proof assistants will
rely on their ability to incorporate computations within deduction in
order to mimic the mathematician when replacing the proof of a
proposition P by the proof of an equivalent proposition P' obtained
from P thanks to possibly complex calculations.

Proof assistants based on the Curry-Howard isomorphism such as Coq
\cite{coqv80} allow to build the proof of a proposition by applying
appropriate proof tactics generating a proof term that can be checked
with respect to the rules of logic.  The proof-checker, also called
the \emph{kernel} of the proof assistant, implements the inference and
deduction rules of the logic on top of a term manipulation layer.
Trusting the kernel is vital since the mathematical correctness of a
proof development relies entirely on the kernel.

The (intuitionist) logic on which Coq is based is the Calculus of
Constructions (CC) of Coquand and Huet \cite{coquand:ic88}, an
impredicative type theory incorporating polymorphism, dependent types
and type constructors. As other logics, CC enjoys a computation
mechanism called cut-elimination, which is nothing but the
$\beta$-reduction rule of the underlying $\lambda$-calculus. But
unlike logics without dependent types, CC enjoys also a powerful
type-checking rule, called {\em conversion}, which incorporates
computations within deduction, making decidability of type-checking
a non-trivial property of the calculus.

The traditional view that computations coincide with
$\beta$-reductions suffers several drawbacks. A methodological one is
that the user must encode other forms of computations as deduction,
which is usually done by using appropriate, complex tactics.  A
practical one is that proofs become much larger than necessary, up to
a point that they cannot be type-checked anymore.  These questions
become extremely important when carrying out complex developments
involving a large amount of computation as the formal proof of the
four colour (now proof-checked) theorem completed by Gonthier and
Werner using Coq~\cite{gonthier:types04}.

The Calculus of Inductive Constructions of Coquand and Paulin was a
first attempt to solve this problem by introducing inductive types and
the associated elimination rules~\cite{coquand89}. The recent versions
of Coq are based on a slight generalization of this
calculus~\cite{gimenez:icalp98}. Besides the $\beta$-reduction rule,
they also include the so-called $\iota$-reductions which are recursors
for terms and types.  While the kernel of CC is extremely compact and
simple enough to make it easily readable -hence trustable-, the kernel
of CIC is much larger and quite complex.  Trusting it would require a
formal proof, which was done once~\cite{barras:thesis}. Updating that
proof for each new release of the system is however unrealistic.  CIC
does not solve our problem, though, since such a simple function as
\emph{reverse} of a \emph{dependent list} cannot be defined in CIC
because $a::l$ and $l::a$, assuming $::$ is list concatenation
and the element $a$ can be coerced to a list of length 1, have
non-convertible types $list(n+1)$ and $list(1+n)$.

A more general attempt was carried out since the early 90's, by adding
user-defined computations as rewrite rules, resulting in the Calculus
of Algebraic Constructions \cite{blanqui:mscs05}. Although
conceptually quite powerful, since CAC captures
CIC~\cite{blanqui:fi05}, this paradigm does not yet fulfill all needs,
because the set of user-defined rewrite rules must satisfy several
strong assumptions. No implementation of CAC has indeed been released
because making type-checking efficient would require compiling the
user-defined rules, a complex task resulting in a kernel too large to
be trusted anymore.

The proof assistant PVS uses a potentially stronger paradigm than Coq
by combining its deduction mechanism\footnote{PVS logic is not based
  on Curry-Howard and proof-checking is not even decidable making both
  frameworks very different and difficult to compare.} with a notion
of computation based on the powerful Shostak's method for combining
decision procedures~\cite{shostak79}, a framework dubbed \emph{little
  proof engines} by Shankar~\cite{shankar:lics02}: the \emph{little
  proof engines} are the decision procedures, required to be convex,
combined by Shostak's algorithm. A given decision procedure encodes a
fixed set of axioms $P$.  But an important advantage of the method is
that the relevant assumptions $A$ present in the context of the proof
are also used by the decision procedure to prove a goal $G$, and
become therefore part of the notion of computation. For example, in
the case where the little proof engines is the congruence closure
algorithm, the fixed set of axioms $P$ is made of the axioms for
equality, $A$ is the set of algebraic ground equalities declared in
the context, while the goal $G$ is an equality $s=t$ between two
ground expressions.  The congruence closure algorithm will then
process $A$ and $s=t$ together in order to decide whether or not $s=t$
follows from $P\cup A$. In the Calculus of Constructions, this proof
must be constructed by a specific tactic called by the user, which
applies the inference rules of CC to the axioms in $P$ and the
assumptions in $A$, and becomes then part of the proof term being
built. Reflexion techniques allow to omit checking this proof term by
proving the decision procedure itself, but the soundness of the entire
mechanism cannot be guaranteed~\cite{corbineau:thesis}.

Two further steps in the direction of integrating decision procedures
into the Calculus of Constructions are Stehr's Open Calculus of
Constructions OCC~\cite{stehr:fi07} and Oury's Extensional Calculus of
Constructions~\cite{oury:tphol05}. Implemented in Maude, OCC allows
for the use of an arbitrary equational theory in conversion. ECC can
be seen as a particular case of OCC in which all provable equalities
can be used in conversion, which can also be achieved by adding the
extensionality and Streicher's axiom~\cite{streicher:ctt98} to CIC,
hence the name of this calculus. Unfortunately, strong normalization
and decidability of type checking are lost in ECC (and OCC), which
shows that we should look for more restrictive extensions. In a
preliminary work, we also designed a new, quite restrictive framework,
the Calculus of Congruent Constructions (CCC), which incorporates the
congruence closure algorithm in CC's conversion~\cite{ccc-draft},
while preserving the good properties of the calculus, including the
decidability of type checking.

\paragraph{Problem.}
The main question investigated in this paper is the incorporation of a
general mechanism calling a decision procedure for solving
conversion-goals in the Calculus of Inductive Constructions which uses
the relevant information available from the current context of the
proof.

\paragraph{Contributions.}
Our main contribution is the definition and the meta-theoretical
investigation of the Calculus of Congruent Inductive Constructions
(CCIC), which incorporates arbitrary \emph{first-order theories} for
which entailment is decidable into deduction via an abstract
conversion rule of the calculus. A major technical innovation of this
work lies in the computation mechanism: goals are sent to the decision
procedure together with the set of user hypotheses available from the
current context. Our main result shows that this extension of CIC does
not compromise its main properties: confluence, strong normalization,
coherence and decidability of proof-checking are all preserved.
Unlike previous calculi, the main difficulty here is confluence, which
led to a complex definition of conversion as a fixpoint. As a
consequence of this definition, decidability of type checking becomes
itself difficult.

Finally, we explain why the new system is still trustable, by leaving
decision procedures \emph{out} of its kernel, assuming that each
procedure delivers a checkable \emph{certificate} which becomes part
of the proof. Certificate checkers become themselves part of the
kernel, but are usually quite small and efficient and can be added one
by one, making this approach a good compromise between CIC and the
aforementioned extensions.

We assume some familiarity with typed lambda
calculi~\cite{barendregt:book92} and the Calculus of Inductive
Constructions.

\newcommand{\da}{\downarrow}

\section{The calculus}
\label{s:calc}

For ease of the presentation, we restrict ourselves to \CCNAT, a
calculus of constructions with a type $\nat$ of natural numbers
generated by its two constructors $\NO$ and $\NS$ and equipped with
its weak recursor $\RecW$. The calculus is also equipped with a
polymorphic equality symbol $\ccieq$ for which we use here a mixfix
notation, writing $t \ccieq_T u$ (or even $t \ccieq u$ when $T$ is not
relevant) instead of $\ccieq T\,t\,u$.

\medskip

Let $\cS = \{ \Prop, \Type, \Extern \}$ the set of \emph{\CCNAT
  sorts}. For $s \in \{ \Prop, \Type \}$, $\cX^s$ denotes a countably
infinite set of \emph{$s$-sorted variables} s.t. $\cX^\Prop \cap
\cX^\Type = \emptyset$. The union $\cX^\Prop \cup \cX^\Type$ will be
written $\cX$. For $x \in \cX$, we write $s_x$ the sort of $x$. Let
$\cA = \{ \Au, \Ar \}$ a set of two constants called
\emph{annotations}, totally ordered by $\Au \prec_\cA \Ar$, where
$\Ar$ stands for \emph{restricted} and $\Au$ for \emph{unrestricted}.
We use $a$ for an arbitrary annotation.

\begin{definition}[Pseudo-terms of \CCNAT]
  We define the \emph{pseudo-terms} of \CCNAT by the grammar rules:
  \begin{center}
    \smallskip
    $\begin{array}{l@{\;}c@{\;}l}
      t, T & := &
      x \in \cX  \vertrel
      s \in \cS  \vertrel
      \nat       \vertrel
      \:\ccieq\: \vertrel
      \NO        \vertrel
      \NS        \vertrel
      \NP        \vertrel
      \Eq(t)     \vertrel
      t\,u       \\[.5em]
      & \vertrel &
      {\lambda [x :^a T] t}     \vertrel
      {\forall (x :^a T) .\, t} \vertrel
      \RecW(t, T)\{t_0,t_S\}
     \end{array}$
     \smallskip
  \end{center}

  We use $\FV(t)$ for the set of free variables of $t$.
\end{definition}

\begin{definition}[Pseudo-contexts of \CCNAT]
  The \emph{typing environments} of \CCNAT are defined as $\Gamma,
  \Delta := [] \vertrel \Gamma,[x :^a T]$ s.t. a variable cannot
  appear twice. We use $\dom(\Gamma)$ for the domain of $\Gamma$ and
  $x\Gamma$ for the type associated to $x$ in $\Gamma$.
\end{definition}

Remark that in our calculus, assumptions stored in the proof context
always come along with an annotation used to control whether they can
be used (in case the annotation is $\Ar$) or not in a conversion goal.
We will later point out why this is necessary.

\begin{definition}[Syntactic classes]
  The pairwise disjoint syntactic classes of $\CCNAT$,
  called \emph{objects} ($\cO$), \emph{predicates} or \emph{types}
  ($\cP$), \emph{kinds} ($\cK$), \emph{externs} ($\cE$) and
  $\Extern$ are defined in Figure~\ref{fig:classes}.

  This enumeration defines a postfixed successor function +1 on
  classes ($\cO+1=\cP$, $\cP+1=\cK$, $\ldots$ $\Delta+1=\bottom$) . We
  also define $\Class(t)=\cD$ if $t \in \cD$ and $\cD \in \{ \cO, \cP,
  \cK, \cE, \Extern \}$ and $\Class(t) = \bottom$ otherwise.
\end{definition}

\begin{myfigure}
$\begin{array}{l@{\;:=\;}l}
      \cO & \cX^\Prop \vertrel \NO \vertrel \NS \vertrel \NP \vertrel
            \cO\,\cO \vertrel \cO\,\cP \vertrel
            [\lambda \cX^\Prop :^a \cP] \cO \vertrel \\[.4em]
          & [\lambda \cX^\Type :^a \cK] \cO \vertrel
            \RecW(\cO,\cdot)\{\cO,\cO\} \\[.5em]
      \cP & \cX^\Type \vertrel \nat \vertrel
            \cP\,\cO \vertrel \cP\,\cP \vertrel
            [\lambda \cX^\Prop :^a \cP] \cP \vertrel
            \,{\ccieq}\, \vertrel \\[.4em]
          & [\lambda \cX^\Type :^a \cK] \cP \vertrel
            (\forall \cX^\Prop :^a \cP) \cP \vertrel
            (\forall \cX^\Type :^a \cK) \cP\\[.5em]
      \cK & \Prop \vertrel
            \cK\,\cO \vertrel \cK\,\cP \vertrel
            [\lambda \cX^\Prop :^a \cP] \cK \vertrel \\[.4em]
          & [\lambda \cX^\Type :^a \cK] \cK \vertrel
            (\forall \cX^\Prop :^a \cP) \cK \vertrel
            (\forall \cX^\Type :^a \cK) \cK\\[.5em]
      \cE & \Type \vertrel
            (\forall \cX^\Prop :^a \cP) \cE \vertrel
            (\forall \cX^\Type :^a \cK) \cE\\[.5em]
      \Extern & \Extern
 \end{array}$
   \caption[ ]{\label{fig:classes} \CCNAT terms classes}
\end{myfigure}

Our typing judgments are classically written $\Gamma \vdash t : T$,
meaning that the \emph{well formed term} $t$ is a proof of the
proposition $T$ under the assumptions in the \emph{well-formed}
environment $\Gamma$.  \emph{Typing rules} are those of CIC restricted
to the single inductive type of natural numbers, with one exception,
\textsc{[Conv]}, based on an equality relation called
\emph{conversion} defined in section~\ref{ss:conv}.

\begin{definition}[Typing]
  Typing rules of \CCNAT are defined in
  Figure~\ref{fig:ccnat-judgement}.
\end{definition}

\begin{myfigure}[hptb]
  \InferRule{Axiom-1}{ }{\vdash \Prop : \Type}
  \hspace{.7cm}
  \InferRule{Axiom-2}{ }{\vdash \Type : \Extern}

  \bigskip

  \InferRule{$\ccieq$-Intro}{ }
            {\vdash \:\ccieq\: : \forall (T :^u \Prop) .\, T \ra T \ra \Prop}

  \bigskip

  \InferRule{Product}
            {\Gamma \vdash T : s_T \quad \Gamma,[x :^a T] \vdash U : s_U}
            {\Gamma \vdash {\forall (x :^a T) .\, U} : s_U}

\bigskip

  \InferRule{Lamda}
            {\Gamma \vdash {\forall (x :^a T) .\, U} : s \quad
             \Gamma,[x :^a T] \vdash u : U}
            {\Gamma \vdash {\lambda [x :^a T] u} : {\forall (x :^a T) .\, U}}

  \bigskip

  \InferRule{Weak}
              {\Gamma \vdash V : s \quad \Gamma \vdash t : T \\
                s \in \{ \Prop, \Type \} \quad x \in \cX^s - \dom(\Gamma)}
              {\Gamma,[x :^a V] \vdash t : T}

  \bigskip

  \InferRule{Var}{x \in \dom(\Gamma) \quad \Gamma \vdash x\Gamma : s_x}
                 {\Gamma \vdash x : x\Gamma}

  \bigskip

  \InferRule{App}
            {\Gamma \vdash t : {\forall (x :^a U) .\, V} \quad
             \Gamma \vdash u : U \\\\
             \Trm{if $a = \Ar$ and $U \rwrefltrans{\beta} t_1 \ccieq_T t_2$
                  with $t_1, t_2 \in \cO$}\\\\
             \Trm{then $t_1 \cnv{\Gamma} t_2$ must hold}}
            {\Gamma \vdash t\,u : V \{ x \mapsto u \}}

  \bigskip

  \InferRule{$\NO$-Intro}{ }{\vdash \NO : \nat}
  \hspace{.7cm}
  \InferRule{$\NS$-Intro}{ }{\vdash \NS : \nat \ra \nat}

  \bigskip

  \InferRule{Nat}{ }{\vdash \nat : \Prop}
  \hspace{.7cm}
  \InferRule{$\NP$-Intro}{ }{\vdash \NP : \nat \ra \nat \ra \nat}

  \bigskip

  \InferRule{Eq-Intro}
            {\Gamma \vdash t_1 : T \quad
             \Gamma \vdash t_2 : T \\\\
             \Gamma \vdash p : \forall (P : T \ra \Prop) .\, P\,t_1 \ra P\,t_2}
            {\Gamma \vdash \Eq(p) : t_1 \ccieq_T t_2}

  \bigskip

  \InferRule{$\iota$-Elim}
            {\Gamma \vdash t : \nat \quad
              \Gamma \vdash Q : \nat \ra \Prop \quad
              \Gamma \vdash f_0 : \nat \\\\
              \Gamma \vdash f_S : {\forall (n :^\Au \nat) .\, Q\,n \ra Q\,(\NS\,n)}}
            {\Gamma \vdash \RecW(t, Q)\{f_0,f_S\} : Q\,t}

  \bigskip

  \InferRule{Conv}
            {\Gamma \vdash t : T \quad \Gamma \vdash T' : s' \quad
             T \cnv{\Gamma} T'}
            {\Gamma \vdash t : T'}

  \medskip
  \caption[ ]{\label{fig:ccnat-judgement} Typing judgment of \CCNAT}
\end{myfigure}

\subsection{Computation by conversion}
\label{ss:conv}
Our calculus has a complex notion of computation reflecting its rich
structure made of three different ingredients, the typed lambda
calculus, the type $\nat$ with its weak recursor and the Presburger
arithmetic.

Our typed lambda calculus comes along with the $\beta$-rule. The
$\eta$-rule raises known technical difficulties,
see~\cite{werner:thesis}.

The type $\nat$ is generated by the two constructors $\NO$ and $\NS$
whose typing rules are given in Figure~\ref{fig:ccnat-judgement}. We
use $\RecW$ for its weak recursor whose typing rule is given in
Figure~\ref{fig:ccnat-judgement} as well. Following CIC's tradition,
we separate their arguments into two groups, using parentheses for the
first two, and curly brackets for the two branches. The computation
rules of $\nat$ are given below:

\begin{definition}[$\iota$-reduction]
  The $\iota$-reduction is defined by the following rewriting system:
  \begin{center}
    $\begin{array}{l@{\;\rw{\iota}\;}l}
      \RecW(\NO,    Q)\{t_0,t_S\} & t_0 \\[.5em]
      \RecW(\NS\,t, Q)\{t_0,t_S\} & t_S \; t \; (\RecW(t, Q)\{t_0,t_S\})\\[.5em]
    \end{array}$
  \end{center}

\vspace{-.5em}
\noindent where $t_0, t_S \in \cO$.
\end{definition}

These rules are going to be part of the conversion $\cnv{\Gamma}$. Of
course, we do not want to type-check terms at each single step of
conversion, we want to type-check only the starting two terms forming
the equality goal in {\textsc [Conv]}. But intermediate terms
could then be non-typable and strong normalization be lost.

The constructors $\NO$ and $\NS$, as well as the additional
first-order constant $\NP$ are \emph{also} used to build up
expressions in the algebraic world of Presburger arithmetic, in which
function symbols have arities. We therefore have two different
possible views of terms of type $\nat$, either as a term of the
calculus of inductive constructions, or as an algebraic term of
Presburger arithmetic. We now define precisely this algebraic world
and explain in detail how to extract algebraic information from
arbitrary terms of $\CCNAT$.

Let $\cT$ be the theory of \emph{Presburger arithmetic} defined on the
signature $\Sigma = \{0, S(\_), \_+\_\}$ and $\cY$ a set of variables
distinct from $\cX$. Note that we syntactically distinguish the
algebraic symbols from the $\CCNAT$ symbols by using a different font
($0$ and $S$ for the algebraic symbols, $\NO$ and $\NS$ for the
constructors).

We write $\cT \vDash F$ if $F$ is a valid formula in $\cT$, and $\cT,
E \vDash F$ for $\cT \vDash E \implies F$.

\begin{definition}[Algebraic terms]
  The set \Alg of \emph{\CCNAT algebraic terms} is the smallest subset
  of $\cO$ s.t.
  \begin{inparaenum}[i)]
    \item $\cX^\Prop \subseteq \Alg$,
    \item $\NO \in \Alg$,
    \item $\forall t \in \CCNAT .\, \NS\,t \in \Alg$,
    \item $\forall t, u \in \CCNAT .\, t \NP u \in \Alg$.
  \end{inparaenum}
\end{definition}

\begin{definition}[Algebraic cap and aliens]
  Given a relation $R$ on $\CCNAT$, let $\cR$ be the smallest
  congruence on $\CCNAT$ containing $R$, and $\pi_R$ a function from
  $\CCNAT$ to $\cY \cup \cX^\Prop$ such that $t \mathrel{\cR} u \iff
  \pi_R(t) = \pi_R(u)$.

  \medskip

  The \emph{algebraic cap of $t$ modulo $R$}, $\fcap_R(t)$, is defined
  by:
  \begin{itemize}
    \item $\fcap_R(\NO) = 0$, $\fcap_R(\NS\,u) = S(\fcap_R(u))$,
      $\fcap_R(u \NP v) = \fcap_R(u) + \fcap_R(v)$,
    \item otherwise, $\fcap_R(t) = t$ if $t \in \cX^*$ and else
      $\pi_R(t)$.
  \end{itemize}
  We call \emph{aliens} the subterms of $t$ abstracted by a variable
  in $\cY$.
\end{definition}

Observe that a term not headed by an algebraic symbol is abstracted by
a variable from our new set of variables $\cY$ in such a way that
$\cR$-equivalent terms are abstracted by the same variable.

We can now glue things together to define \emph{conversion}.

\begin{definition}[Conversion relation]
The family $\{\cnv{\Gamma}\}_{\Gamma}$ of $\Gamma$-\emph{conversions}
  is defined by the rules of Figure~\ref{fig:conversion-relation}.
\end{definition}

{\afterpage{
\begin{myfigure}[hptb]
  \InferRule{$\beta\iota$}
    {t \rwequiv{\beta\iota} u}
    {t \cnv{\Gamma} u}
  \hspace{.7cm}
  \InferRule{Eq}
    {[x :^\Ar T] \in \Gamma \quad
     T \rwrefltrans{\beta} t_1 \mathrel{\ccieq} t_2 \quad
     t_1, t_2 \in \cO}
    {t_1 \cnv{\Gamma} t_2}

  \bigskip

  \InferRule{Ded}
    {t_1, t_2 \in \cO \\\\
     \cT, \{ \fcap_{\cnv{\Gamma}}(u_1) = \fcap_{\cnv{\Gamma}}(u_2) \vertrel
              u_1 \cnv{\Gamma} u_2\} \vDash
         \fcap_{\cnv{\Gamma}}(t_1) = \fcap_{\cnv{\Gamma}}(t_2)}
    {t_1 \cnv{\Gamma} t_2}

  \bigskip

  \InferRule{Sym}{t \cnv{\Gamma} u}{u \cnv{\Gamma} t}
  \hspace{.7cm}
  \InferRule{Trans}
            {t \cnv{\Gamma} u \quad u \cnv{\Gamma} v}
            {t \cnv{\Gamma} v}

  \bigskip

  \InferRule{\CCNAT-Eq}{t \cnv{\Gamma} u}{\Eq(t) \cnv{\Gamma} \Eq(u)}
  \hspace{.7cm}
  \InferRule{App}
            {t_1 \cnv{\Gamma} t_2 \quad u_1 \cnv{\Gamma} u_2}
            {t_1\,u_1 \cnv{\Gamma} t_2\,u_2}
  \bigskip

  \InferRule{Prod}
            {T \cnv{\Gamma} U \quad t \cnv{\Gamma,[x :^a T]} u \quad b \preceq a}
            {{\forall (x :^b T) .\, t} \cnv{\Gamma} {\forall (x :^b U) .\, u}}
  \hspace{.7cm}
  \InferRule{Lam}
            {T \cnv{\Gamma} U \quad t \cnv{\Gamma,[x :^a T]} u \quad b \preceq a}
            {{\lambda [x :^b T] t} \cnv{\Gamma} {\lambda [x :^b U] u}}

  \bigskip

  \InferRule{Elim-$\cW$}
            {t \cnv{\Gamma} u \quad P \cnv{\Gamma} Q \quad
             t_0 \cnv{\Gamma} u_0 \quad t_S \cnv{\Gamma} u_S}
            {\RecW(t, P)\{t_0, t_S\} \cnv{\Gamma} \RecW(u, Q)\{u_0, u_S\}}

  \medskip
  \caption[ ]{\label{fig:conversion-relation} Conversion relation
    $\cnv{\Gamma}$}
\end{myfigure}}}

This definition is technically complex.

Being a congruence, $\cnv{\Gamma}$ includes congruence rules. However,
all these rules are not quite congruence rules since crossing a
binder increases the current context $\Gamma$ by the new assumption
made inside the scope of the binding construct, resulting in a family
of congruences. More questions are raised by the three different kinds
of basic conversions.

First, $\cnv{\Gamma}$ includes the rules $\rw{\beta}$ and $\rw{\iota}$
of $\CCNAT$.  Unlike the beta rule, $\rw{\iota}$ interacts with
first-order rewriting, and therefore the \textsc{Conv} rule of
Figure~\ref{fig:ccnat-judgement} cannot be expressed by $T
\rwequiv{\beta\iota} \cnv{\Gamma} \rwequiv{\beta\iota} T'$ as one
would expect.

Second, $\cnv{\Gamma}$ includes the relevant assumptions grabbed from
the context, this is rule \textsc{Eq}. These assumptions must be of
the form $[x :^\Ar T]$, with the appropriate annotation $r$, and $T$
must be an equality assumption or otherwise \emph{reduce} to an
equality assumption. Note that we use only $\rw{\beta}$ here.  Using
$\cnv{\Gamma}$ recursively instead is indeed an equivalent formulation
under our assumptions. Without annotations, $\CCNAT$ does not enjoy
subject reduction. Generating appropriate annotations is discussed in
section~\ref{s:discussion}.

Third, with rule \textsc{[Ded]}, we can also generate new assumptions
by using Presburger arithmetic. This rule here uses the property that
two algebraic terms are equivalent in $\cnv{\Gamma}$ if their caps
relative to $\cnv{\Gamma}$ are equivalent in $\cnv{\Gamma}$ (the
converse being false).  This is so because the abstraction function
$\pi_{\cnv{\Gamma}}$ abstracts equivalent aliens by the same variable
taken from $\cY$. It is therefore the case that deductions on caps
made in Presburger arithmetic can be lifted to deductions on arbitrary
terms via the abstraction function. As a consequence, the two
definitions of the abstraction function $\pi_{\cnv{\Gamma}}$ and of
the congruence $\cnv{\Gamma}$ are mutually inductive: our conversion
relation is defined as a least fixpoint.

\subsection{Two simple examples}

\paragraph{More automation - smaller proofs.}
We start with a simple example illustrating how the equalities
extracted from a context $\Gamma$ can be use to deduce new equalities
in $\cnv{\Gamma}$.

\begin{center}
$\begin{array}{@{}l@{\:}l@{\:}l@{}}
  \Gamma & = & {[x\,y\,t :^\Au \nat],[f :^\Au \nat \ra \nat]},\\
         & &   {[p_1 :^\Ar t \ccieq 2]},
               {[p_2 :^\Ar f\,(x \NP 3) \ccieq x \NP 2]},\\
         & &   {[p_3 :^\Ar f\,(y \NP t) \NP 2 \ccieq y]},
               {[p_4 :^\Ar y \NP 1 \ccieq x \NP 2]}
\end{array}$
\end{center}

From $p_1$ and $p_4$ (extracted from the context by \textsc{[Eq]}),
\textsc{[Ded]} will deduce that $y \NP t \cnv{\Gamma} x \NP 3$, and by
congruence, $f\,(y \NP t) \cnv{\Gamma} f\,(x \NP 3)$. Therefore,
$\pi_{\cnv{\Gamma}}$ will abstract $f(x \NP 3)$ and $f(y \NP t)$ by
the same variable $z$, resulting in two new equations available for
\textsc{[Ded]}: $z=x+2$ and $z+2=y$.  Now, $z=x+2$, $z+2=y$ and
$y+1=x+2$ form a set of unsatisfiable equations and we deduce $0
\cnv{\Gamma} 1$ by the \textsc{Ded} rule: contradiction has been
obtained. This shows that we can easily carry out a proof by
contradiction in $\cT$.

\paragraph{More typable terms.}
We continue with a second example showing that the new calculus can
type more terms than CIC.  For the sake of this example we assume that
the calculus is extended by dependent lists on natural numbers. We
denote by $\dlist$ (of type $\nat \ra \Prop$) the type of dependent
lists and by $\Lnil$ (of type $\dlist\,\NO$) and $\Lcons$ (of type
$\forall (n : \nat) .\, \dlist\,n \ra \nat \ra \dlist\,(\NS\,n)$) the
lists constructors. We also add a weak recursor $\RecLW$ such that,
given $P : \forall (n : \nat) .\, \dlist\,n \ra \Prop$, $l_0 :
P\,\NO\,\Lnil$ and $l_S : \forall (n : \nat)(l : \dlist\,n) .\,
P\,n\,l \ra \forall (x : \nat) .\, P\,(\NS\,n)\,(\Lcons\,n\,x\,l)$,
then $\RecLW(l,P)\{l_0,l_S\}$ has type $P\,n\,l$ for any list $l$ of
type $\dlist\,n$.

\smallskip

Assume now given a dependent $\reverse$ function (of type $\forall (n
: \nat) .\, \dlist\,n \ra \dlist\,n$) and the list concatenation
function $@$ (of type $\forall (n\,n': \nat), \dlist\,n \ra \dlist\,n'
\ra \dlist\,(n \NP n')$). We can simply express that a list $l$ is a
palindrome: $l$ is a palindrome if $\reverse\,l \ccieq l$.

\smallskip

Suppose now that one wants to prove that palindromes are closed under
substitution of letters by palindromes. To make it easier, we will
simply consider a particular case: the list $l_1 l_2 l_2 l_1$ is a
palindrome if $l_1$ and $l_2$ are palindromes. The proof sketch is
simple: it suffices to apply as many times as needed the lemma
$\reverse(l l') = \reverse(l') @ \reverse(l)$~$(*)$. What can be quite
surprising is that Lemma~$(*)$ is rejected by Coq. Indeed, if $l$ and
$l'$ are of length $n$ and $n'$, it is easy to check that $\reverse(l
l')$ is of type $\dlist\,(n \NP n')$ and $\reverse(l') :: \reverse(l)$
of type $\dlist\,(n' \NP n)$ which are clearly not
$\beta\iota$-convertible. This is not true in our system: $n \NP n'$
will of course be convertible to $n' \NP n$ and lemma $(*)$ is
therefore well-formed. Proving the more general property needs of
course an additional induction on natural numbers to apply lemma $(*)$
the appropriate number of times, which can of course be carried out in
our system.

\smallskip

Note that, although possible, writing a $\reverse$ function for
dependent lists in Coq is not that simple. Indeed, a direct inductive
definition of $\reverse$ will define $\reverse(\Lcons\,n\,a\,l)$, of
type $\dlist\,(1 \NP n)$, as $\reverse(l) \mathrel{@} a$, of type
$\dlist\,(n \NP 1)$. Coq will reject such a definition since
$\dlist\,(1 \NP n)$ and $\dlist\,(n \NP 1)$ are not convertible.
Figure~\ref{f:coqreverse} shows how $\reverse$ can be defined in Coq.

\begin{myfigure}
\begin{small}
\begin{flushleft}
\texttt{Coq~{<}~Definition~reverse:~forall~(n:~nat),~(list~n)~-{>}~(list~n)~.}\\
\texttt{Coq~{<}~~~assert~(reverse\_acc~:~forall~(n~m~:~nat),}\\
\texttt{Coq~{<}~~~~~~~~~~~~~list~n~-{>}~list~m~-{>}~list~(m+n))~.}\\
\texttt{Coq~{<}~~~refine~(fix~reverse\_acc~(n~m~:~nat)~(from~:~list~n)~(to~:~list~m)}\\
\texttt{Coq~{<}~~~~~~~~~~~~~\{struct~from\}~:~list~(m+n)~:=~\_)~.}\\
\texttt{Coq~{<}~~~destruct~from~as~[~|~n'~v~rest~]~.}\\
\texttt{Coq~{<}~~~~~rewrite~{<}-~plus\_n\_0\_transparent;~exact~to~.}\\
\texttt{Coq~{<}~~~~~rewrite~{<}-~plus\_n\_Sm\_transparent;}\\
\texttt{Coq~{<}~~~~~~~exact~(reverse\_acc~n'~(S~m)~rest~(cons~\_~v~to))~.}\\
\texttt{Coq~{<}~~~intros~n~l~.~exact~(reverse\_acc~\_~\_~l~nil)~.}\\
\texttt{Coq~{<}~Defined~.}\\
\end{flushleft}
\end{small}

\caption[ ]{\label{f:coqreverse} $\reverse$ function is Coq}
\end{myfigure}

\section{Metatheorical properties}
\label{s:theory}

Most basic properties of Pure Type Systems (see \cite{blanqui:thesis})
are not too difficult. Those using substitution instances are more
delicate.  They rely on the annotations decorating the abstractions
and products which were introduced for that purpose.

\subsection{Stability by substitution}
\label{ss:propconv}

Assume that $\Gamma$ is a typing environment of the form $\Gamma_1,[ p
:^\Ar a \ccieq b ],\Gamma_2$ ($a$ and $b$ being two variables of type
$\nat$ in $\Gamma$).  The stability by substitution claims that if we
have a typing derivation $\Gamma \vdash t : T$, then we can substitute
$p$ by a term $P$ (of type $a \ccieq b$ under $\Gamma_1$) in this
derivation and obtain a proof of $\Gamma_1,\Gamma_2\theta \vdash
t\theta : T\theta$, where $\theta$ is the substitution $\{ p \mapsto P
\}$. This property can easily be proved for Pure Type Systems as soon
as the conversion relation is itself stable by substitution. In our
example one can easily check that $a \cnv{\Gamma} b$, but $a
\cnv{\Gamma_1,\Gamma_2\theta} b$ will not hold in general: the
assumption $a \ccieq b$ has been inlined and thus is no more
extractable by the conversion relation.  As a result, we need to
strengthen the formulation of stability by substitution:

\begin{lemma}
  Let $\Gamma = \Gamma_1,[z :^a W],\Gamma_2$ and assume that
  \begin{inparaenum}[i)]
  \item $T \cnv{\Gamma} T'$,
  \item if $a = \Ar$ and $W \rwrefltrans{\beta} t_1 \ccieq t_2$
    then $t_1 \cnv{\Gamma_1} t_2$.
  \end{inparaenum}
  Then, $T\theta \cnv{\Delta} T'\theta$ where $\theta = \{ z \mapsto w
  \}$ and $\Delta = \Gamma_1,\Gamma_2\theta$
\end{lemma}

\begin{corollary}[Stability by substitution]
  \label{l:substitutivity}
  Let $\Gamma = \Gamma_1,[z :^a W],\Gamma_2$ and assume that
  \begin{inparaenum}[i)]
    \item $T \cnv{\Gamma} T'$
    \item if $a = \Ar$ and $W \rwrefltrans{\beta} t_1 \ccieq t_2$ then
      $t_1 \cnv{\Gamma_1} t_2$.
  \end{inparaenum}
  Then, $\Delta \vdash t\theta : T\theta$ where $\theta = \{ z \mapsto
  w \}$, $\Gamma_1 \vdash w : W$ and $\Delta =
  \Gamma_1,\Gamma_2\theta$.
\end{corollary}

As usual, the substitutivity lemma is to be used in the proof of
subject reduction (for~$\rw{\beta\iota}$) to come later. Because it
requires a specific typing property for the equality assumptions
annotated by $\Ar$, we need to ensure this property in the application
case of the coming subject reduction proof. This is indeed the origin
of the similar condition arising in the typing rule \textsc{[App]}.

\subsection{Conversion as rewriting}

We now turn conversion into a rewriting relation in order to prove
that our system is logically consistent by analyzing a proof in normal
form of $\forall (x :^\Au \Prop) .\, x$. The notion of a normal proof
is of course more complicated than in CIC, since we must account for
the congruence $\cnv{\Gamma}$ associated with an arbitrary context
$\Gamma$. The difficulty is that the set of equalities assumed in a
given environment $\Gamma$ together with the axioms of the theory
$\cT$ may be inconsistent, making all first-order terms equal in
$\cnv{\Gamma}$ which could break strong normalization of our rewriting
relation.

\begin{definition}[$\cT$-consistent environment]
  A typing environment $\Gamma$ is \emph{$\cT$-consistent} if there
  exist two terms $t,u\in\cO$ s.t. $\neg(t \cnv{\Gamma} u)$.
\end{definition}

\begin{lemma}
  \label{l:O-S-inconvertible}
  If $\Gamma$ is $\cT$-consistent then $\neg(\NO \cnv{\Gamma} \NS\,t)$
  for any term $t$.
\end{lemma}

\begin{definition}[Weak conversion]
  We inductively define a family of \emph{weak conversion relations}
  $\{ \cnvbl{\Gamma} \}_\Gamma$ as the smallest congruent relation
  s.t. $t \cnvbl{\Gamma} u$ if $\cT, \Eq(\Gamma) \vDash
  \fcap_\emptyset(t) = \fcap_\emptyset(u)$, where $\Eq(\Gamma) = \{
  \fcap_\emptyset(w_1) = \fcap_\emptyset(w_2) \vertrel w_1, w_2 \in
  \cO, [x :^\Ar w_1 \ccieq w_2] \in \Gamma \}$.
\end{definition}

\begin{definition}
We inductively define a family $\{\rw{\Gamma}\}_\Gamma$ of rewriting
relations modulo weak-conversion as the smallest rewriting relations
satisfying the rules of Figure~\ref{fig:conv-as-rewriting}.
\end{definition}

The first rule shows that rewriting is modulo weak conversion in a
consistent environment. The second equates all object terms when the
environment is inconsistent, replacing them by the new constant
$\bullet$. The others are as expected.

\begin{myfigure}
  \InferRule{Rw-Mod}
            {\Trm{$\Gamma$ is $\cT$-consistent} \quad
          t \cnvbl{\Gamma} t' \rw{\Gamma} u' \cnvbl{\Gamma} u}
            {t \rw{\Gamma} u}

  \medskip

  \InferRule{Rw-$\bullet$}
            {\Trm{$\Gamma$ is $\cT$-inconsistent} \quad t \in \cO \quad t \ne \bullet}
            {t \rw{\Gamma} \bullet}

  \medskip

  \InferRule{Rw-$\beta\iota$}
            {t \rw{\beta\iota} u}{t \rw{\Gamma} u}
  \hspace{.7cm}
  \InferRule{Rw-Fwd}
            {t \rw{\Delta} u \quad \Gamma \rw{\beta} \Delta}
            {t \rw{\Gamma} u}

  \medskip

  \InferRule{W-$\forall$}
            {t \rw{\Gamma,[x :^a T]} u \quad b \preceq a}
            {\forall (x :^b T) .\, t \rw{\Gamma} \forall (x :^b T) .\, u}

  \medskip

  \InferRule{W-$\lambda$}
              {t \rw{\Gamma,[x :^a T]} u \quad b \preceq a}
              {\lambda [x :^b T] .\, t \rw{\Gamma} \lambda [x :^b T] .\, u}

 \medskip
 \caption[ ]{\label{fig:conv-as-rewriting} Conversion as a rewriting system}
\end{myfigure}

\begin{lemma}
  \begin{enumerate}
  \item The rewriting relation $\rw{\Gamma}$ is confluent.
  \item If $t \cnv{\Gamma} u$ then $t \rwequiv{\Gamma} u$.
  \item If $t \rwequiv{\Gamma} u$ with $\bullet \not\in t$ and
    $\bullet \not\in u$ then $t \cnv{\Gamma} u$.
  \item If $\Gamma \vdash t : T$ with $\Gamma$ $\cT$-consistent and $t
    \cnvbl{\Gamma} u$, then $\Gamma \vdash u : T$.
  \end{enumerate}
\end{lemma}

\begin{lemma}
  \label{l:sn}
  If $\Gamma \vdash t : T$ and $t \rw{\Gamma} u$ with $\bullet \not\in
  u$, then $\Gamma \vdash u : T$.
\end{lemma}
\begin{proof}
  The proof is standard, by induction on the type derivation of the
  left-hand side. The interesting case is when a $\beta$-reduction
  applies to the top of a term of the form $(\lambda [x:^a U] v)~w$
  and the typing rule is \textsc{[App]}: we then conclude by using
  Lemma~\ref{l:substitutivity}. Note that the side condition of rule
  \textsc{[App]} provides us with the property needed for using
  Lemma~\ref{l:substitutivity}.
\end{proof}

\begin{lemma}
  The rewriting relation $\rw{\Gamma}$ is strongly normalizing for
  well formed terms.
\end{lemma}
\begin{proof}
  The proof is a direct application of proof
  irrelevance~\cite{barthe98}, because $\cnv{\Gamma}$ is a
  congruence generated by equalities between object terms, apart
  from beta-reduction. What makes this true is that $\RecW$ is a
  weak recursor, working at the object level. Including strong
  elimination rules invalidates this argument. \qed
\end{proof}

We finally conclude  that $\CCNAT$ is consistent:

\begin{theorem}
  \label{l:consistency}
  There is no proof of $\:$ $\vdash t : \forall (x :^\Au \Prop) .\, x$.
\end{theorem}

\begin{proof}
  Assume that $\vdash t : \NO \ccieq \NS\,\NO$ where $t$ is
  $\rw{\Gamma}$-normal.  Since $\NO \ccieq \NS\,\NO$ is not
  convertible to a sort, $t$ cannot be equal to $\nat$, or a sort, or
  a product. Since $t$ is necessarily closed, $t$ is not a variable.
  Moreover, $t$ cannot be of the form $\RecW(u, Q)\{ t_0, t_S\}$ since
  $t$ is closed and in $\rw{\iota}$-normal form.

  If $t$ is an application, it is necessarily of the form $c\,\vec{u}$
  with $c \in \{\NO, \NS, \NP, \ccieq\}$. By using inversion it
  suffices to check that in all these cases, $t$ has a type $T$ which
  is not convertible to $\NO \ccieq \NS\,\NO$.

  If $t = \Eq(u)$, then $t$ has type $u \ccieq u$ with $u$ of type
  $\nat$ and $u \ccieq u$ convertible to $\NO \ccieq \NS\,\NO$. Thus
  $\NO \cnv{[]} \NS\,\NO$, and $\cT \vDash 0 = 1$, which is
  impossible. \qed
\end{proof}

\subsection{Decidability of type checking}

\begin{theorem}
  Type checking of $\CCNAT$ is decidable.
\end{theorem}

Decidability of type checking needs two ingredients. First-of-all,
eliminating \textsc{[Conv]}, which is non-structural, by incorporating
it to \textsc{[App]}. This is classical, and it is easy to prove
decidability of the transformed set of rules for type-checking,
assuming $\cnv{\Gamma}$ is decidable.

\medskip

Deciding $\cnv{\Gamma}$ is more complex. We cannot use the rewrite
system $\rw{\Gamma}$ for that purpose since the first two rules use
the $\cT$-consistency of $\Gamma$ as a prerequisite. We use instead a
saturation based algorithm. The method resembles very much the one
used for combining first-order decision procedures operating on
disjoint alphabets~\cite{mss:jsc89,baader92cade}. Basic ingredients
are: purification of formulas (here equations) by abstracting aliens
by new variables; deriving new equalities among variables by using the
appropriate decision procedure for pure formulas; propagating these
new equalities to the other formulas.

\section{Conclusion and discussion}
\label{s:discussion}

$\CCNAT$ is an extension of CIC (restricted to the weak elimination
rules of the inductive type $\nat$) by a fragment of Presburger
arithmetic (without the natural strict order $\mathbb{N}$) in which
conversion incorporates Presburger arithmetic, $\beta$-reduction and
higher-order primitive recursion into a single mechanism. We now
discuss in more details how this can be generalized to full CIC, how
this can be used in practice, how useful that is, and whether the
obtained kernel is trustable.

\paragraph{Relevance.}
Our second example shows very clearly the expressivity of our calculus
with respect to CIC. However, what is done here by a typing rule could
be done alternatively in CIC by a tactic. Besides, if one wants to
avoid building a proof term which can be quite large and slow down the
type-checker, it is possible to prove the tactic and then use a
reflexion mechanism in order to avoid type-checking the proof each
time the tactic is called. In both cases, however, the user must call
the tactic explicitly. In our approach, this is completely
transparent, and would remain transparent in case of a succession of
uses of the decision procedure separated by eliminations, since
conversion incorporates both, or in case of different decision
procedures called successively.

\paragraph{Extension to CIC.}
Building decision procedures in a type-theoretic framework is not that
easy.  The main difficulty lies in the adequate definition of the
congruence $\cnv{\Gamma}$. Once the definition is obtained, carrying
out the technical development is not too difficult in the case of the
pure Calculus of Constructions (the congruence becomes quite simpler
in this case), difficult in the present case of $\CCNAT$ (because of
the presence of the weak recursors for $\nat$), no more difficult when
other decidable theories are introduced such as lists with their
associated recursors, but much harder when including strong
elimination rules which interact with the first-order theories.  In
this case, it is necessary to block the congruence below the strong
recursor in order to avoid lifting an incoherence from the object
level to the predicate level, which would immediately yield
paradoxes~\cite{strub:thesis}.

\paragraph{Annotations restriction}
One may wonder how annotations can be handled in practice. As seen,
annotations are used to forbid \emph{inlining} (when a $\beta$-redex
is contracted) of equational assumptions which are used by conversion.
This could be seen as a restriction since our calculus, in order to
avoid the creation of problematic $\beta$-redexes, forbids in most
cases applications of symbols of type $\forall (p :^\Ar t \ccieq u)
.\, T$.

This restriction can be removed by using the notion of \emph{opaque
  definitions} (as opposed to \emph{transparent definitions}) of Coq
which allows the user to define symbols that the system cannot inline.
In most cases, definitions having a computational behavior (like
$\NP$) are transparent whereas definitions representing lemmas (like
the associativity of $\NP$) are opaque. This convention is used in the
standard library of Coq.

Returning to our previous example, if the user needs to prove a lemma
of the form $\forall (p :^\Ar t \ccieq u) .\, T$, he or she should
declare it as an opaque definition $P := \lambda [p :^\Ar t \ccieq u]
q$. The application of $P$ to a term $v$ should then be allowed: the
term $P\,v$ cannot reduce to $q \{ p \mapsto v \}$. Of course, if $P$
is defined transparently, the application $P\,v$ has to be forbidden.

Moreover, this gives us a simple heuristic to automatically tag
products and abstractions: $\Ar$ annotation should by used by default
when the user is defining an opaque symbol, whereas $\Au$ annotation
should be used everywhere else.

\paragraph{Arbitrary decision procedures.}
So far, we have considered only decidable equational theories. But it
is well-known that a decidable theory can always be transformed into a
decidable equational theory over the type Bool of truth values
equipped with its usual operations. This is so because of the
decidability assumption.

\paragraph{Type levels equalities.}
One may wonder whether the conversion relation of $\CCNAT$ could use
type level equalities (or hypotheses of the form $P \lra Q$). The
answer seems to be negative: extracting type levels equalities breaks
subject reduction and $\beta$-strong normalization
(see~\cite{oury:tphol05}), two properties needed for the decidability
of our calculus.

\paragraph{Trusting the kernel.}
Decision procedures require complex coding. It took a lot of time to
get a correct tactic for Presburger arithmetic in Coq. Including a
tactic into the kernel of the system is therefore unrealistic, unless
it is itself proved correct with a trustable proof assistant. On the
other hand, most decision procedures can provide a \emph{certificate}
that is quite compact and can be verified by a
\emph{certificate-checker} which is usually small, and easy to write
and read, and is therefore a trustable piece of code.  The reason is
that the procedure \emph{searches} for a proof while the
certificate-checker \emph{verifies} that the certificate is correct. A
certificate checker looks indeed like a proof-checker. It is then easy
to modify the conversion rule so as to output a certificate each time
a decision procedure is used. The kernel of $\CCNAT$ should therefore
include a certificate-checker for Presburger arithmetic. In case of
CCIC with several decision procedures, the kernel would include one
proof-checker for each decision procedure. Besides, the process is
incremental: the procedures and the associated proof-checkers can be
included one by one, because decision procedures for different
inductive types operate on disjoint vocabularies, hence can be
combined~\cite{mss:jsc89,baader92cade}.

\medskip

An implementation of CCIC has started and should be available soon as
a prototype in a version without certificate generation and checking.

\bibliographystyle{plain}

\begin{thebibliography}{10}

\bibitem{baader92cade}
F.~Baader and K.~Schulz.
\newblock Unification in the union of disjoint equational theories: Combining
  decision procedures.
\newblock In Deepak Kapur, editor, {\em Proc. 11th Int. Conf. on Automated
  Deduction, Saratoga Springs, NY, LNAI 607}, 1992.

\bibitem{barendregt:book92}
H.~Barendregt.
\newblock {\em Lambda calculi with types}, volume~2 of {\em Handbook of logic
  in computer science}.
\newblock Oxford University Press, 1992.

\bibitem{barras:thesis}
B.~Barras.
\newblock {\em Auto-validation d'un système de preuves avec familles
  inductives}.
\newblock PhD thesis, Université de Paris~VII, 1999.

\bibitem{barthe98}
G.~Barthe.
\newblock The relevance of proof irrelevance.
\newblock In {\em Proc. 24th Int. Coll. on Automata, Languages and Programming,
  LNCS 1443}, LNCS, 1998.

\bibitem{blanqui:thesis}
F.~Blanqui.
\newblock {\em Type Theory and Rewriting}.
\newblock PhD thesis, Université de Paris~XI, Orsay, France, 2001.

\bibitem{blanqui:mscs05}
F.~Blanqui.
\newblock Definitions by rewriting in the calculus of constructions.
\newblock {\em Mathematical Structures in Computer Science}, 15(1):37--92,
  2005.
\newblock Journal version of LICS'01.

\bibitem{blanqui:fi05}
F.~Blanqui.
\newblock Inductive types in the calculus of algebraic constructions.
\newblock {\em Fundamenta Informaticae}, 65(1-2):61--86, 2005.
\newblock Journal version of TLCA'03.

\bibitem{ccc-draft}
F.~Blanqui, J.-P. Jouannaud, and P.-Y. Strub.
\newblock A {C}alculus of {C}ongruent {C}onstructions.
\newblock Unpublished draft, 2005.

\bibitem{coqv80}
Coq-Development-Team.
\newblock {\em The {C}oq Proof Assistant Reference Manual - Version 8.0}.
\newblock INRIA, INRIA Rocquencourt, France, 2004.
\newblock At URL \url{http://coq.inria.fr/}.

\bibitem{coquand:ic88}
T.~Coquand and G.~Huet.
\newblock The {C}alculus of {C}onstructions.
\newblock {\em Information and {C}omputation}, 76(2-3):95--120, 1988.

\bibitem{coquand89}
Th. Coquand and C.~Paulin-Mohring.
\newblock Inductively defined types.
\newblock In Martin-Löf and G.~Mints, editors, {\em Colog'-88, International
  Conference on Computer Logic}, volume 417 of {\em LNCS}, pages 50--66.
  Springer-Verlag, 1990.

\bibitem{corbineau:thesis}
P.~Corbineau.
\newblock {\em Démonstration automatique en Théorie des Types}.
\newblock PhD thesis, University of Paris~IX, 2005.

\bibitem{gimenez:icalp98}
E.~Giménez.
\newblock Structural recursive definitions in type theory.
\newblock In {\em Proceedings of ICALP'98}, volume 1443 of {\em LNCS}, pages
  397--408, July 1998.

\bibitem{gonthier:types04}
G.~Gonthier.
\newblock The four color theorem in coq.
\newblock In {\em TYPES 2004 International Workshop}, 2004.

\bibitem{streicher:ctt98}
M.~Hofmann and T.~Streicher.
\newblock The groupoid interpretation of type theory.
\newblock In {\em Twenty-five years of constructive type theory}, volume~36 of
  {\em Oxford Logic Guides}, pages 83--111. Oxford University Press, 1998.

\bibitem{oury:tphol05}
Nicolas Oury.
\newblock Extensionality in the calculus of constructions.
\newblock In Joe Hurd and Thomas~F. Melham, editors, {\em TPHOLs}, volume 3603
  of {\em Lecture Notes in Computer Science}, pages 278--293. Springer, 2005.

\bibitem{mss:jsc89}
M.~Schmidt-Schau{\ss}.
\newblock Unification in a combination of arbitrary disjoint equational
  theories.
\newblock {\em J. Symbolic Computation}, 8:51--99, 1989.
\newblock Special issue on Unification.

\bibitem{shankar:lics02}
N.~Shankar.
\newblock Little engines of proof.
\newblock In G.~Plotkin, editor, {\em Proceedings of the Seventeenth Annual
  IEEE Symp. on Logic in Computer Science, {LICS} 2002}. IEEE Computer Society
  Press, 2002.
\newblock Invited Talk.

\bibitem{shostak79}
R.~E. Shostak.
\newblock An efficient decision procedure for arithmetic with function symbols.
\newblock {\em J. of the Association for Computing Machinery}, 26(2):351--360,
  1979.

\bibitem{stehr:fi07}
M.O. Stehr.
\newblock The {O}pen {C}alculus of {C}onstructions: An equational type theory
  with dependent types for programming, specification, and interactive theorem
  proving (part {I} and {II}).
\newblock {\em To appear in Fundamenta Informaticae}, 2007.

\bibitem{strub:thesis}
P.-Y. Strub.
\newblock {\em Type Theory and Decision Procedures}.
\newblock PhD thesis, École Polytechnique, Palaiseau, France, Work in
  progress.

\bibitem{werner:thesis}
B.~Werner.
\newblock {\em Une Théorie des Constructions Inductives}.
\newblock PhD thesis, University of Paris~VII, 1994.

\end{thebibliography}

\end{document}